\documentclass[journal]{IEEEtran}

\usepackage{cite}

\ifCLASSINFOpdf

\usepackage{mhchem}
\usepackage{siunitx}
\usepackage{color}
\usepackage{bm}%
\usepackage[colorlinks=false,linkcolor=blue]{hyperref}
 \usepackage[pdftex]{graphicx}
 \usepackage{float}
 
 \graphicspath{{graphics/pdf/}}

\begin{document}

\title{Structural and magnetic properties of sputter deposited Mn-Fe-Ga thin films}

\author{
\IEEEauthorblockN{Alessia Niesen\(^{1}\), Christian Sterwerf\(^{1}\), Manuel Glas\(^{1}\), Jan-Michael Schmalhorst\(^{1}\) and G\"unter Reiss\(^{1}\)}\\[2mm]
\IEEEauthorblockA{\small{\(^{1}\)Center for Spinelectronic Materials and Devices, Physics Department, Bielefeld University, Germany}} \\[1mm]
}
\date{\today}%


\maketitle

\begin{abstract}
We investigated structural and magnetic properties of sputter deposited \ce{Mn}-\ce{Fe}-\ce{Ga} compounds. The crystallinity of the \ce{Mn}-\ce{Fe}-\ce{Ga} thin films was confirmed using x-ray diffraction. X-ray reflection and atomic force microscopy measurements were utilized to investigate the surface properties, roughness, thickness and density of the deposited \ce{Mn}-\ce{Fe}-\ce{Ga}. Depending on the stoichiometry, as well as the used substrates (\ce{SrTiO3} (001) and \ce{MgO} (001)) or buffer layer (\ce{TiN}) the \ce{Mn}-\ce{Fe}-\ce{Ga} crystallized in the cubic or the tetragonally distorted phase. Anomalous Hall effect and alternating gradient magnetometry measurements confirmed strong perpendicular magnetocrystalline anisotropy. Hard magnetic behavior was reached by tuning the composition. TiN buffered \ce{Mn_{2.7}Fe_{0.3}Ga} revealed sharper switching of the magnetization compared to the unbuffered layers. 

\end{abstract}


\IEEEpeerreviewmaketitle

\section{Introduction}

Ferromagnetic, fully spin polarized materials (half-metals) found a lot of interest in the recent years due to their possible application in spintronic devices as nonvolatile memories\cite{Wolf:2001fu} and field programmable logic devices.\cite{Reiss2007,Thomas2006} To maintain the thermal stability at shrinking device sizes, an out-of-plane (oop) oriented magnetization of the material is advantageous. Therefore investigation of perpendicularly magnetized Heusler compounds has found attraction. High spin polarization and out-of-plane magnetization direction was predicted for the \ce{Mn_{3-x}Ga} (\(0.15 \leq x \leq 2\)) compound.\cite{Balke:2007eb, Winterlik:2008fj, Wurmehl:2006db} The transition from the cubic \(\mathrm{D0_{3}}\) into the tetragonal \(\mathrm{D0_{22}}\) phase takes place at temperatures above 500\si{\degreeCelsius}.\cite{Glas:2013we} Hence a variation of the deposition temperature is an important criterium in terms of the crystallographic properties. The tetragonally distortion could also be obtained by shifting the Fermi energy at the Van Hove singularity in one of the spin-channels, which is reached by tuning the material with an additional element.\cite{Winterlik:2012cp} \ce{Mn}-\ce{Fe}-\ce{Ga} is one possible material, which is calculated to be \SI{95}{\%} spin polarized at the Fermi level for the cubic phase (\ce{Mn_{2}Fe_{1}Ga}).\cite{Wollmann:2014ke} The predicted low total magnetization \(M=1.03\, \mu_{B}\) \cite{Wollmann:2014ke} and high Curie temperature \(T_{C}=\SI{550}{\kelvin}\) (lowest measured value for \ce{Mn_{1.4}Fe_{1.6}Ga}) makes this material interesting to serve as an electrode in magnetic tunnel junctions (MTJ's).\cite{Felser:kh} The replacement of \ce{Mn} atoms by \ce{Fe} leads to an enhancement of the magnetic moment. The measured magnetic moment of pure \ce{Mn}-\ce{Ga} (prepared by arc melting) is \(1\, \mu_{B}\). The \ce{Fe} - rich \ce{Fe_{2}Mn_{1}Ga} showed the highest magnetic moment of \(3.5\, \mu_{B}\).\cite{Felser:kh} 

The tunable magnetic behavior makes this material interesting for investigations and a promissing candidate for applications. 
In this work, we focused on the preparation and investigation of the crystallographic, structural and magnetic properties of the ternary \ce{Mn}-\ce{Fe}-\ce{Ga} compound thin films. The influence of different stoichiometries, substrates and deposition temperatures on the material properties was analyzed.
With regard to the preparation of magnetic tunnel junctions and the aim to increase their applicability, additionally a \ce{TiN} seed-layer was used. Sputter deposited \ce{TiN} is a material with low electrical resistivity (\SI{16}{\micro\ohm \cm}) and a surface roughness below \SI{1}{\nano\m}.\cite{Magnus:2011wl,Krockenberger:2012bo} Thus it provides a good electrical connection to the MTJ. High thermal stability (melting point \SI{2950}{\degreeCelsius} \cite{Pritschow:07}) is another advantage, which prevents chemical reactions of \ce{TiN} with the on top deposited material. The lattice constant of \ce{TiN} (fcc structure) is \SI{4.24}{\angstrom} and therefore suitable for various Heusler compounds. It was already shown, that \ce{TiN} is a suitable seed-layer for \ce{Mn_{3-x}Ga} and \ce{Co_{2}FeAl}.\cite{Niesen2015}

\section{Experimental}
\ce{Mn}-\ce{Fe}-\ce{Ga} thin films (\SI{40}{\nano\m} thickness) were prepared in an ultra-high-vacuum (UHV) sputtering system with a base pressure below \SI{5e-10}{\milli\bar}. DC magnetron co-sputtering from a pure \ce{Mn}, \ce{Fe} and a \ce{Mn45Ga55} composite target was used to prepare the samples. The Ar pressure was set to \SI{1.7e-3}{\milli\bar}. The amount of \ce{Mn}, in the \ce{Mn_{y}Fe_{x}Ga} compound, was varied in the range of \(1.5 \leq y \leq 3\) and the amount of \ce{Fe} in the range of \(0.3 \leq x \leq 1\). Deposition temperatures from \SI{190}{\degreeCelsius} to \SI{595}{\degreeCelsius} were chosen in order to achieve crystalline growth and the tetragonally distorted phase of \ce{Mn}-\ce{Fe}-\ce{Ga}. \ce{MgO} (100) (\(a_{\ce{MgO}} = \SI{4.21}{\angstrom}\)) and \ce{SrTiO3} (\ce{STO}) (100) (\(a_{\ce{STO}} = \SI{3.91}{\angstrom}\)) single crystalline substrates were utilized. Additionally \ce{TiN} buffered \ce{Mn}-\ce{Fe}-\ce{Ga} thin films on \ce{MgO} and \ce{STO} substrates were prepared. The \ce{TiN} layers (\SI{30}{\nano\m}) were deposited using reactive sputtering in an Ar and N atmosphere which results in stoichiometric \ce{Ti1N1} thin films. During the sputtering process a \ce{N} flow of \SI{2}{sccm} and an \ce{Ar} flow of \SI{20}{sccm} was used, leading to a deposition pressure of \SI{1.6e-3}{\milli\bar}. The stoichiometry of \ce{TiN} was verified via density, resistivity and x-ray absorption spectroscopy measurements. Additionally the superconductance of \ce{TiN} was proved. Further information on the \ce{TiN} seed-layer is given in \cite{Niesen2015}. On top of the Heusler compound a \SI{2}{\nano\m} thick \ce{MgO} layer was deposited to prevent the surface from contaminations.

\section{Characterization of the Mn-Fe-Ga compound}

\begin{figure}[t!]%
\includegraphics[width=3.5in]{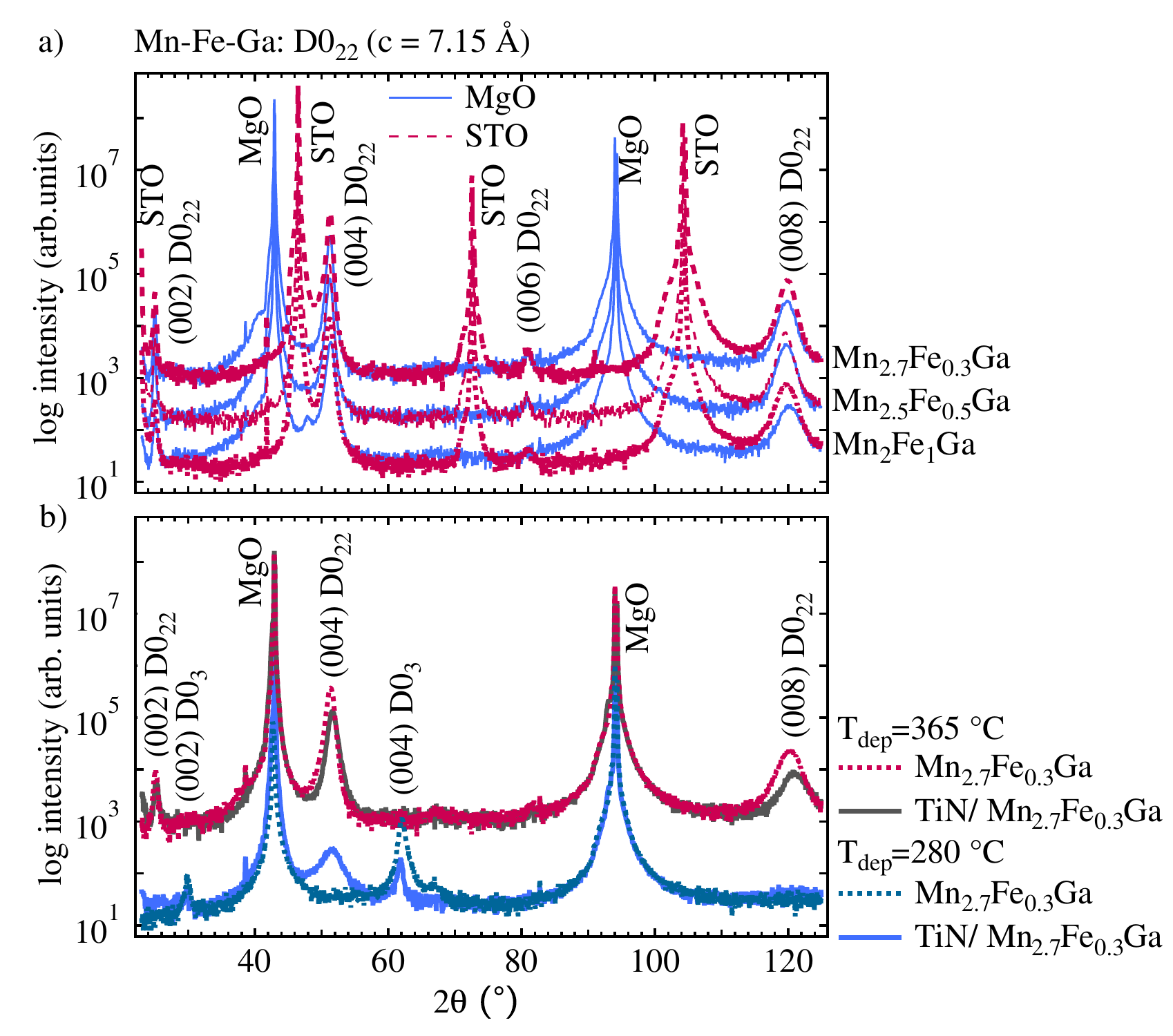}%
\caption{a) XRD patterns of the \ce{Mn_{3-x}Fe_{x}Ga} compound deposited at \SI{450}{\degreeCelsius} on \ce{MgO} (blue curves) and \ce{SrTiO3} (dashed red curves) substrates. b) XRD patterns of \ce{Mn_{2.7}Fe_{0.3}Ga} deposited at \SI{365}{\degreeCelsius} and \SI{280}{\degreeCelsius} with and without a  \ce{TiN} buffer layer.}%
\label{fig:XRD_MnFeGa}%
\end{figure}%

\begin{figure}[t!]%
\centering
\includegraphics[width=3in]{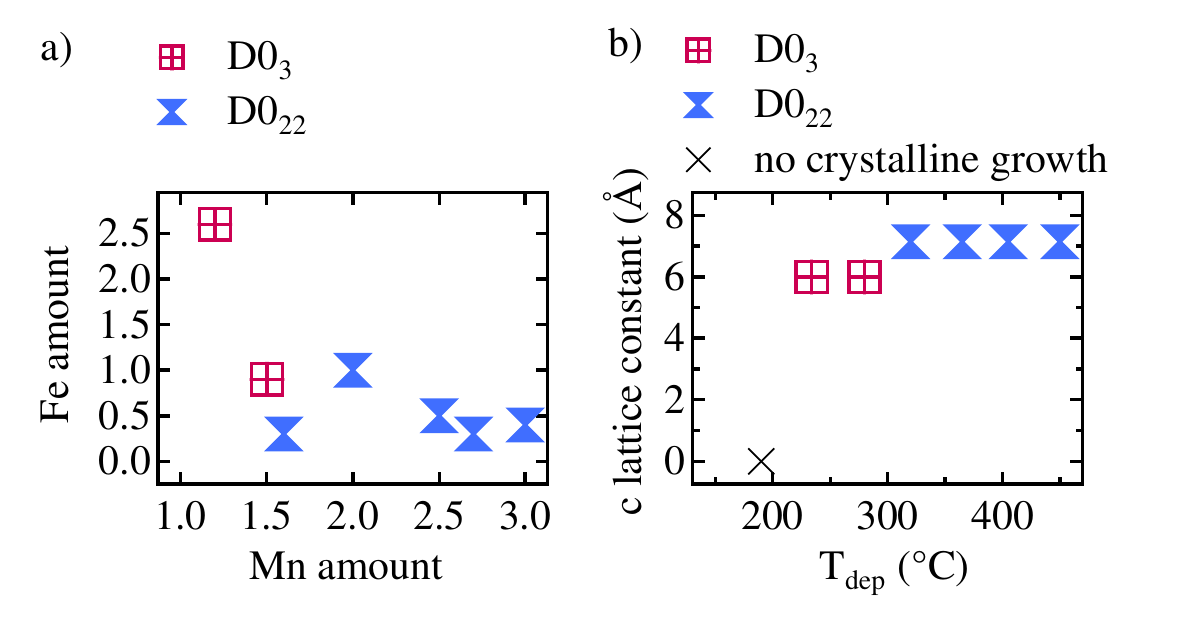}%
\caption{a) Crystal structure of the \ce{Mn}-\ce{Fe}-\ce{Ga} compounds deposited on \ce{MgO} substrates in dependence on the \ce{Mn} and \ce{Fe} amount. b) Dependence of the crystal structure of \ce{Mn_{2.7}Fe_{0.3}Ga} on the deposition temperature.}%
\label{fig:stoch}%
\end{figure}%

\subsection{Crystal structure}

\begin{figure}%
\includegraphics[width=3.5in]{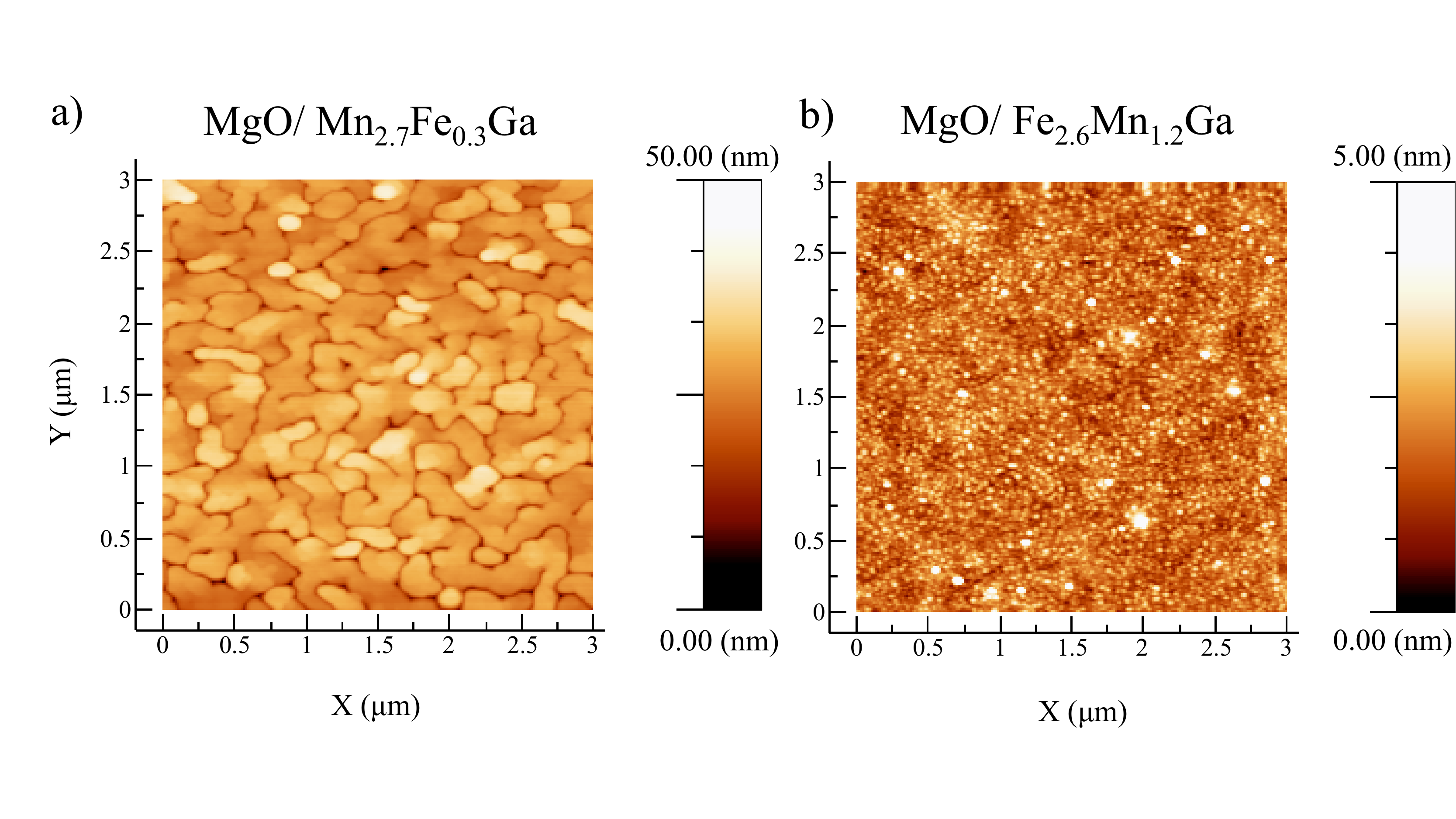}%
\vspace{-1cm}
\caption{AFM images of a) \ce{Mn_{2.7}Fe_{0.3}Ga} (tetragonally distorted phase) and b) \ce{Fe_{2.6}Mn_{1.2}Ga} (cubic phase) thin films deposited on MgO (100) substrates at \SI{450}{\degreeCelsius}.}
\label{fig:AFM_Stoch}
\includegraphics[width=3.5in]{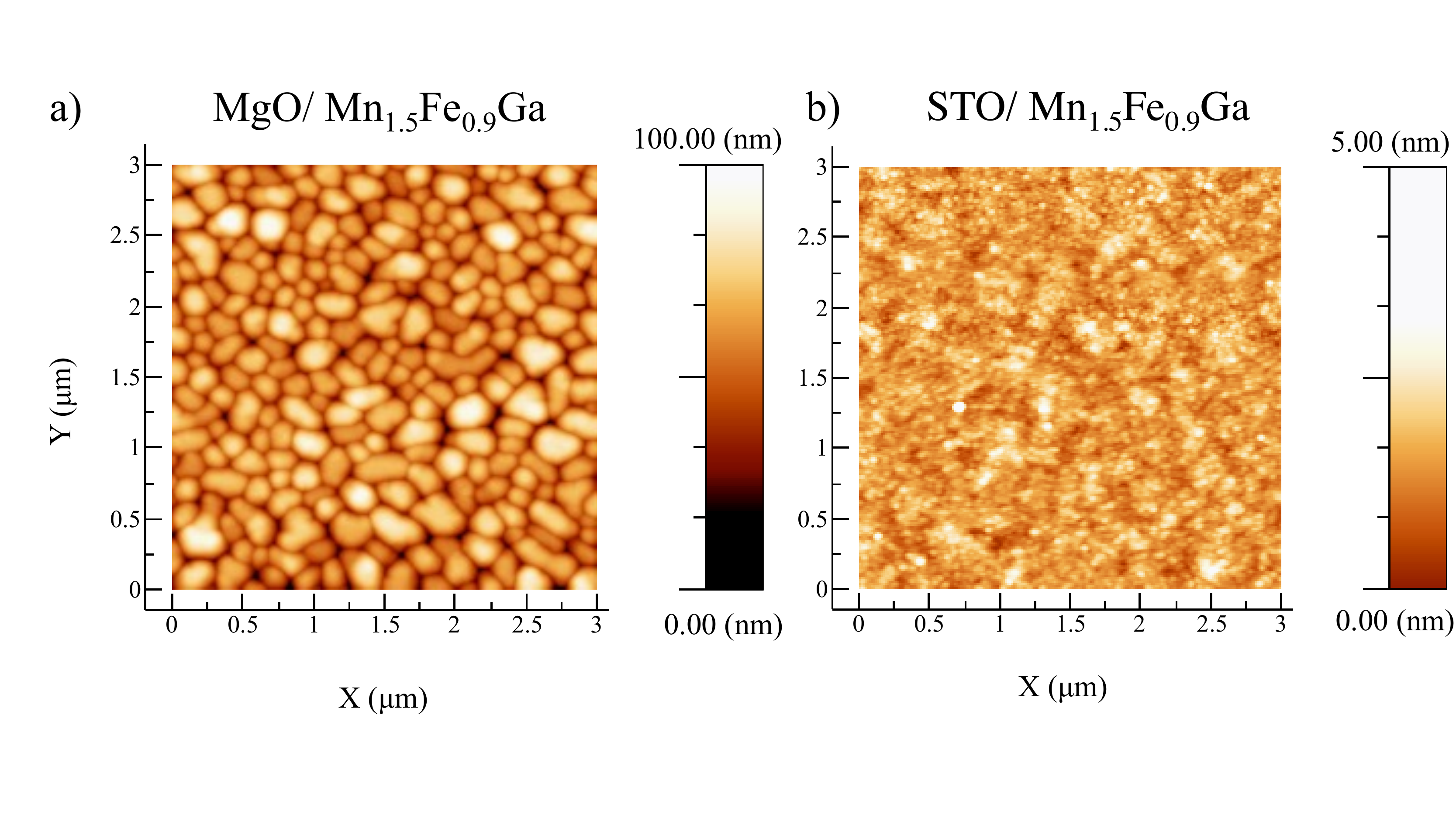}%
\vspace{-1cm}
\caption{AFM images of \ce{Mn_{1.6}Fe_{0.9}Ga} (tetragonally distorted) deposited on a) MgO (100) and b) STO (100) substrates at \SI{550}{\degreeCelsius}.}
\label{fig:AFM_MgO_STO}
\includegraphics[width=3.5in]{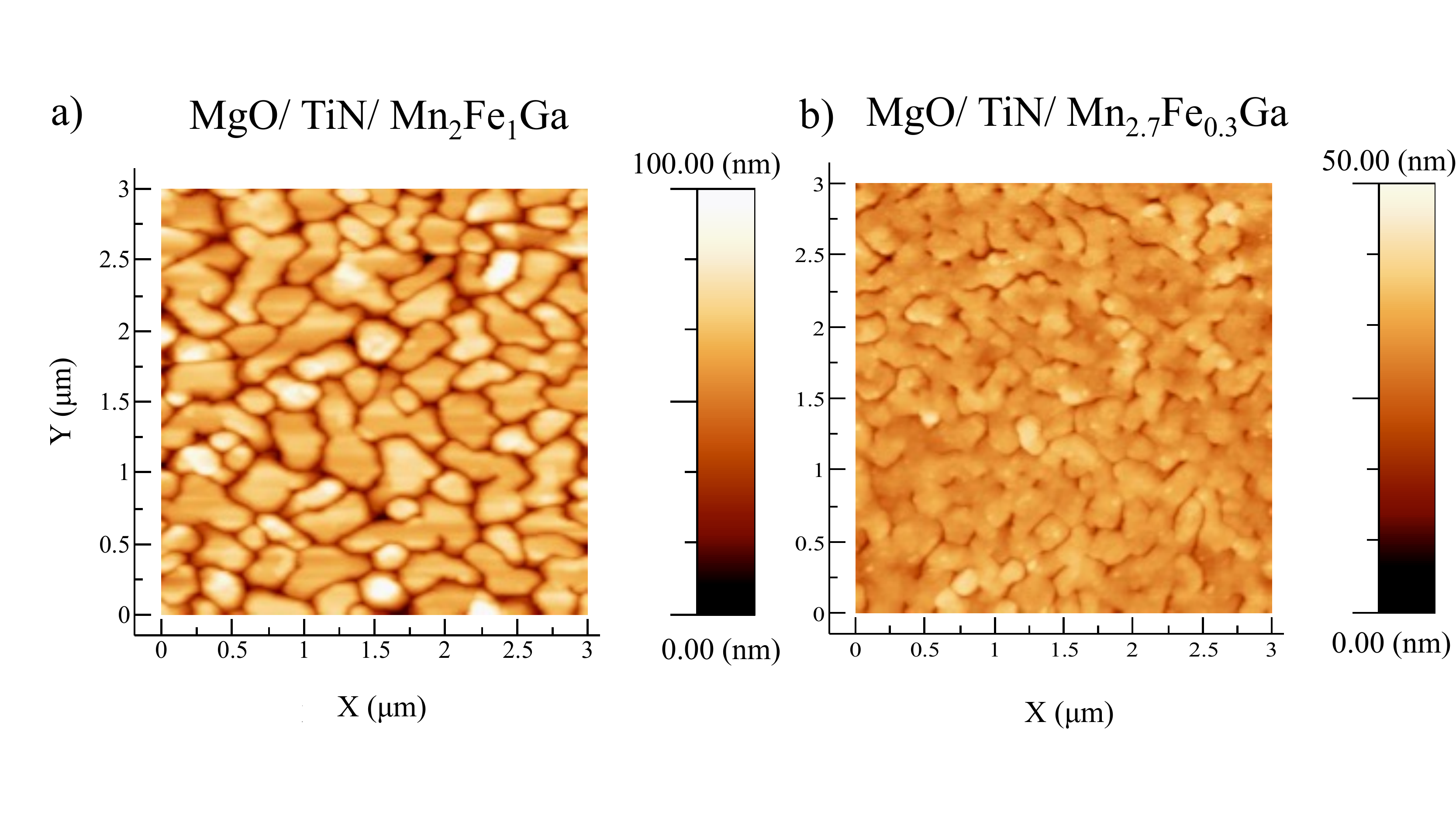}
\vspace{-1cm}
\caption{AFM images of TiN buffered a) \ce{Mn_{2}Fe_{1}Ga} and b) \ce{Mn_{2.7}Fe_{0.3}Ga} deposited on MgO (100) at \SI{450}{\degreeCelsius}.}
\label{fig:AFM_TiN_MFG}
\end{figure}%

Crystallographic and structural properties of the \ce{Mn}-\ce{Fe}-\ce{Ga} thin films were determined via x-ray diffraction (XRD) and x-ray reflection (XRR) measurements using a XRD Philips X'Pert Pro diffractometer (\ce{Cu} anode). The lowest roughness values (\(\leq\) \SI{1}{\nano\m}) combined with the highest crystallinity were measured on samples deposited at temperatures of \SIlist{407;450}{\degreeCelsius}. All samples with the \ce{Mn_{3-x}Fe_{x}Ga} (\(0.3 \leq x \leq 1\)) stoichiometry crystallize in the D0\(_{22}\) tetragonally distorted phase (Fig. \ref{fig:XRD_MnFeGa} and Fig. \ref{fig:stoch}) which is in good agreement with the previously reported data.\cite{Felser:kh} The determined in-plane and oop lattice constants are \(a\) = \num{3.9}\(\pm 0.01\)\,\SI{}{\angstrom} and \(c\) = \num{7.15}\(\pm 0.04\)\,\SI{}{\angstrom} for \ce{Mn}-\ce{Fe}-\ce{Ga} deposited on both substrate types, leading to a \(c/a\) ratio of \num{1.8}. The substrate type obviously does not influence the crystallographic phase of the \ce{Mn_{3-x}Fe_{x}Ga}. Therefore this stoichiometry is stabilized in the tetragonally distorted phase. Increasing the \ce{Fe} and decreasing the \ce{Mn} amount leads to a formation of the cubic phase (D0\(_{3}\)) (\(c\) = \SI{6}{\angstrom}) on \ce{MgO} substrates (Fig. \ref{fig:stoch}). On \ce{STO} each composition results in a formation of the tetragonally distorted phase, due to the low lattice mismatch of \(0.5 \% \) with the in-plane lattice constant. The dependence of the crystal structure on the deposition temperature was investigated for the \ce{Mn_{2.7}Fe_{0.3}Ga} compound, which showes the tetragonally distortion on each substrate type, as well as on the \ce{TiN} buffer layer. The transition from the cubic D0\(_{3}\) (\(a=c=\)\,\SI{6}{\angstrom}) into the tetragonally distorted D0\(_{22}\) phase takes place at a deposition temperature of \SI{320}{\degreeCelsius} (Fig. \ref{fig:stoch}). The density of the samples was determined via XRR measurements and also revealed a dependence on the crystallographic phase. The density of the tetragonally distorted \ce{Mn}-\ce{Fe}-\ce{Ga} is \num{7.1}\(\pm 0.5\)\,\SI{}{\gram/\centi\m^{3}} and \num{6}\(\pm 0.5\)\,\SI{}{\gram/\centi\m^{3}} for the cubic phase.\newline Deposition on a \ce{TiN} buffer layer (\(a_{\ce{TiN}} = \SI{4.24}{\angstrom}\)) leads to a mixture of the cubic D0\(_{3}\) and the D0\(_{22}\) phase, depending on the \ce{Mn}-\ce{Fe}-\ce{Ga} composition and the deposition temperature. The tetragonally distorted phase for the \ce{Mn_{2.7}Fe_{0.3}Ga}, on a \ce{TiN} seed-layer, appears already at a deposition temperature of \SI{280}{\degreeCelsius} (Fig. \ref{fig:XRD_MnFeGa}), which is a lower temperature compared to the unbuffered sample. The seed-layer obviously influences the crystalline growth of this compound leading to a lower deposition temperature at which the tetragonally distorted phase is formed. We already observed this behavior for the \ce{Mn}-\ce{Ga} compound.\cite{Niesen2015} \ce{TiN} buffered \ce{Mn_{2.5}Fe_{0.5}Ga} forms a mixture of the cubic and the tetragonally distorted phase. In case of \ce{TiN} buffered \ce{Mn_{2}Fe_{1}Ga}, only the cubic phase was formed.

\begin{figure}%
\centering
\includegraphics[width=3in]{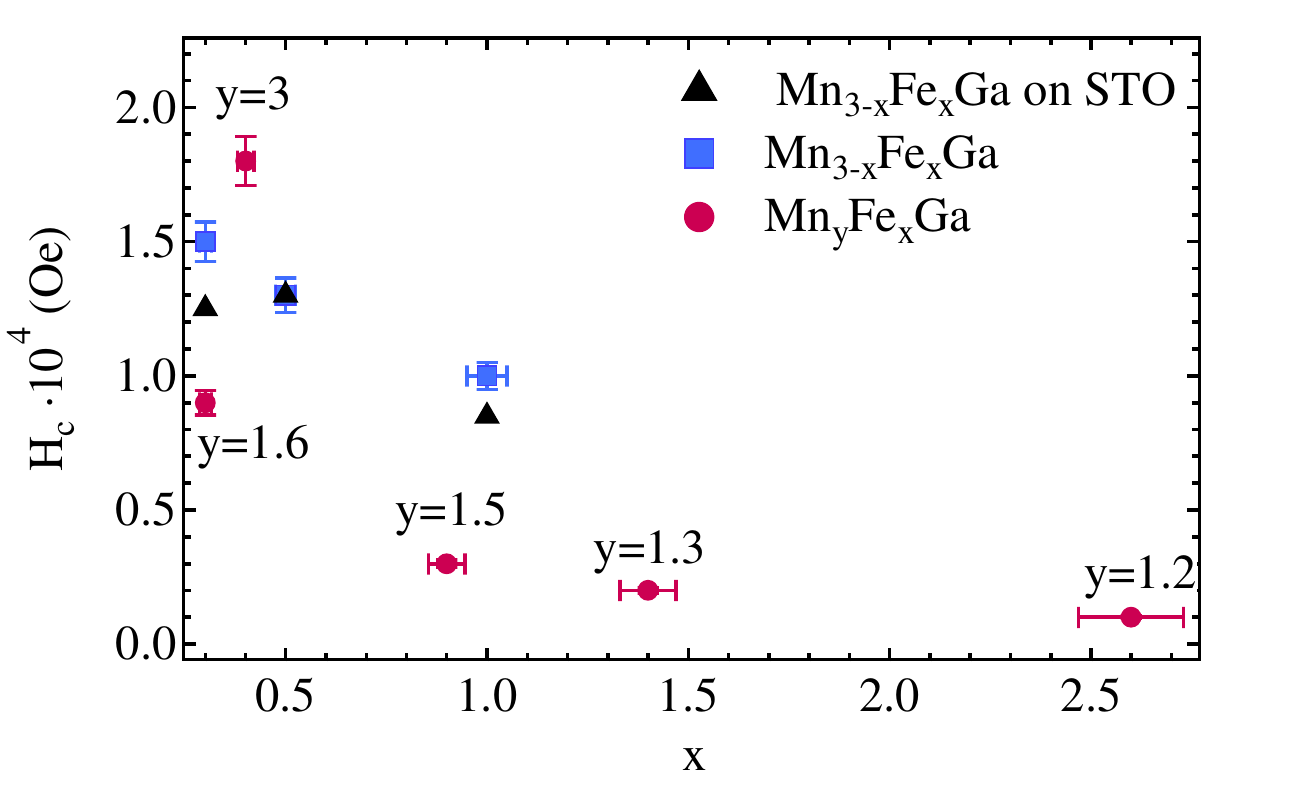}%
\caption{Coercivity dependence on the stoichiometry of \ce{Mn}-\ce{Fe}-\ce{Ga} deposited on \ce{MgO} and \ce{STO} substrates. The blue squaress and black triangles give the coercivity values for \ce{Mn_{3-x}Fe_{x}Ga}. The red circles point out the values for the \ce{Mn_{y}Fe_{x}Ga} stoichiometries (y gives the values for the amount of \ce{Mn}).}%
\label{fig:Hc_vs_Fe_T}%
\end{figure}%

\subsection{Surface properties}

Atomic force microscopy (AFM) was carried out to investigate the surface topography of the samples. 
Fig. \ref{fig:AFM_Stoch} shows a comparison of  (a) the tetragonally distorted \ce{Mn_{2.7}Fe_{0.3}Ga} and (b) the cubic \ce{Fe_{2.6}Mn_{1.2}Ga} layer deposited on \ce{MgO} at \SI{450}{\degreeCelsius}. The \ce{Mn_{2.7}Fe_{0.3}Ga} forms \SI{200}{\nano\m} - \SI{400}{\nano\m} broad grains with steep grain boundaries. The measured rms roughness value is \num{3.4}\(\pm 0.05\)\,\SI{}{\nano\m}. The cubic \ce{Fe_{2.6}Mn_{1.2}Ga} compound forms small grains and a smooth surface (roughness = \num{0.8}\(\pm 0.05\)\,\SI{}{\nano\m}). \ce{Mn}-\ce{Fe}-\ce{Ga} with lower \ce{Mn} ratio shows island growth (Fig. \ref{fig:AFM_MgO_STO} a). The obtained roughness value is \num{17}\(\pm 0.5\)\,\SI{}{\nano\m}. The equivalent film (temperature and stoichiometry) on \ce{STO} revealed no island growth and low roughness of \num{0.43}\(\pm 0.05\)\,\SI{}{\nano\m} (Fig. \ref{fig:AFM_MgO_STO} b), which can be attributed to the low lattice mismatch with the in-plane lattice constant of the tetragonally distorted \ce{Mn}-\ce{Fe}-\ce{Ga}. 

We indicate that the applied roughness analysis is leading to an overestimation of the roughness for samples, which consist of big grains with steep grain boundaries. The structure of \ce{Mn}-\ce{Fe}-\ce{Ga} deposited on a \ce{TiN} seed-layer also showed a strong dependence on the stoichiometry. As previously mentioned, \ce{TiN} buffered \ce{Mn_{2}Fe_{1}Ga} crystallizes in the cubic D0\(_{3}\) phase. Due to the high lattice mismatch of the cubic phase with the lattice constant of \ce{TiN} (\(8.3\%\)) island growth and high roughness (\num{14.25}\(\pm 0.05\)\,\SI{}{\nano\m}) appears (Fig. \ref{fig:AFM_TiN_MFG} a). The \ce{TiN} buffered \ce{Mn_{2.7}Fe_{0.3}Ga}, crystallized in the D0\(_{22}\) phase, shows a similar morphology and roughness value as the unbuffered sample (Fig. \ref{fig:AFM_Stoch} a) and (Fig. \ref{fig:AFM_TiN_MFG} b). Without a \ce{TiN} buffer layer we determined a roughness value of \num{3.4}\(\pm 0.05\)\,\SI{}{\nano\m} (see above) compared to \num{3.54}\(\pm 0.05\)\,\SI{}{\nano\m} with a \ce{TiN} buffer.

\subsection{Magnetic properties}

\begin{table}
\centering
\includegraphics[width=3.2in]{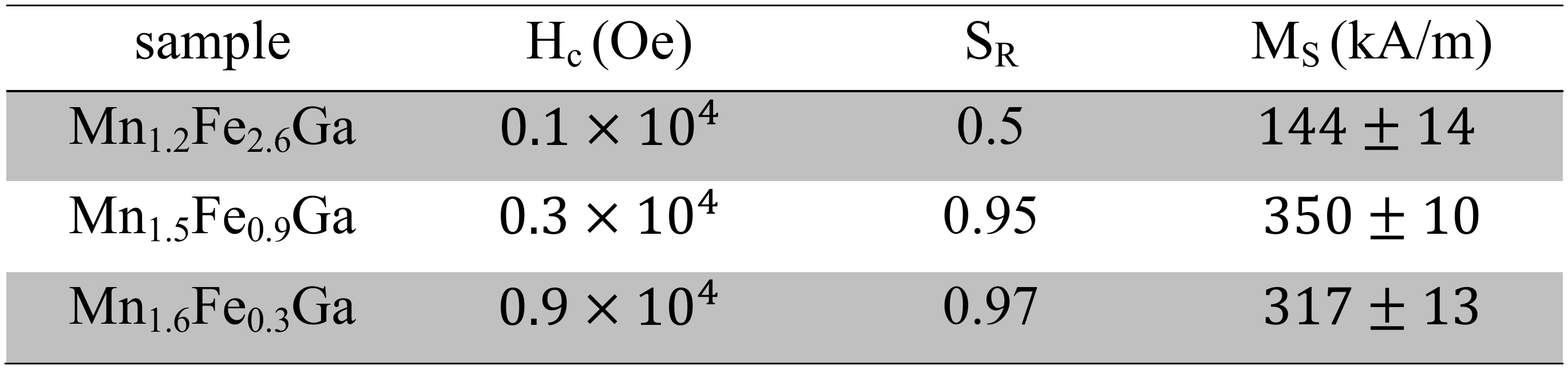}
\vspace{-3.5cm}
\caption{Coercivity \(H_{c}\), squareness ratio \(S_{R}\) and saturation magnetization \(M_{S}\) values of \ce{Mn}-\ce{Fe}-\ce{Ga} deposited on \ce{STO} determined via AHE and AGM measurements. }
\label{TAB1:Hc}
\end{table}

\begin{figure}[t!]%
\centering
\includegraphics[width=3.5in]{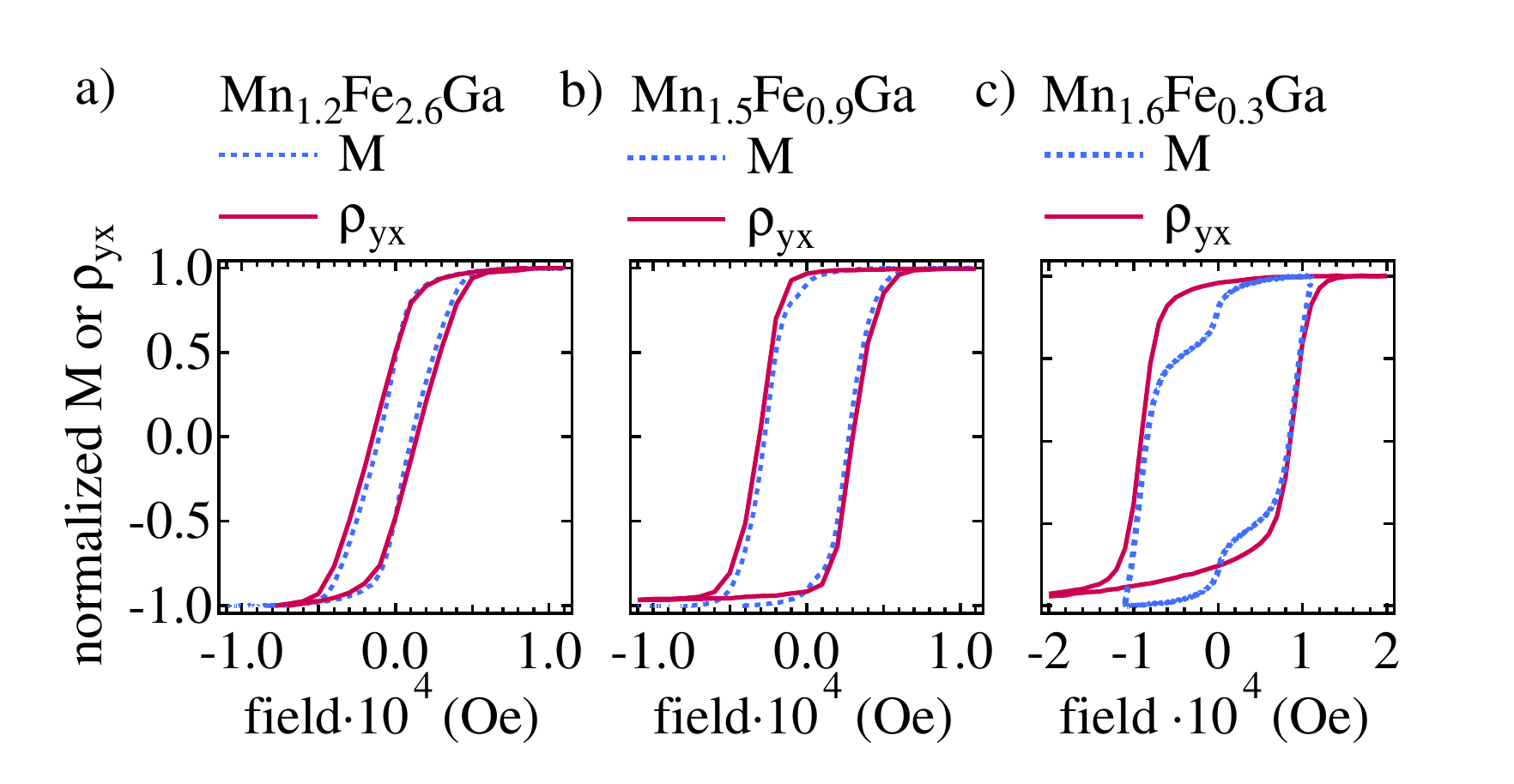}%
\caption{oop measured AHE (red) and AGM (dashed blue) hysteresis curves for a) \ce{Mn_{1.2}Fe_{2.6}Ga}, b) \ce{Mn_{1.5}Fe_{0.9}Ga} and c) \ce{Mn_{1.6}Fe_{0.3}Ga} thin films deposited on \ce{STO}. The measured magnetization \(M\) and Hall voltage values \(U_{H}\) are normalized for a better comparison. 
}%
\label{fig:AHE}%
\end{figure}
\begin{figure}[]%
\centering
\includegraphics[width=3.5in]{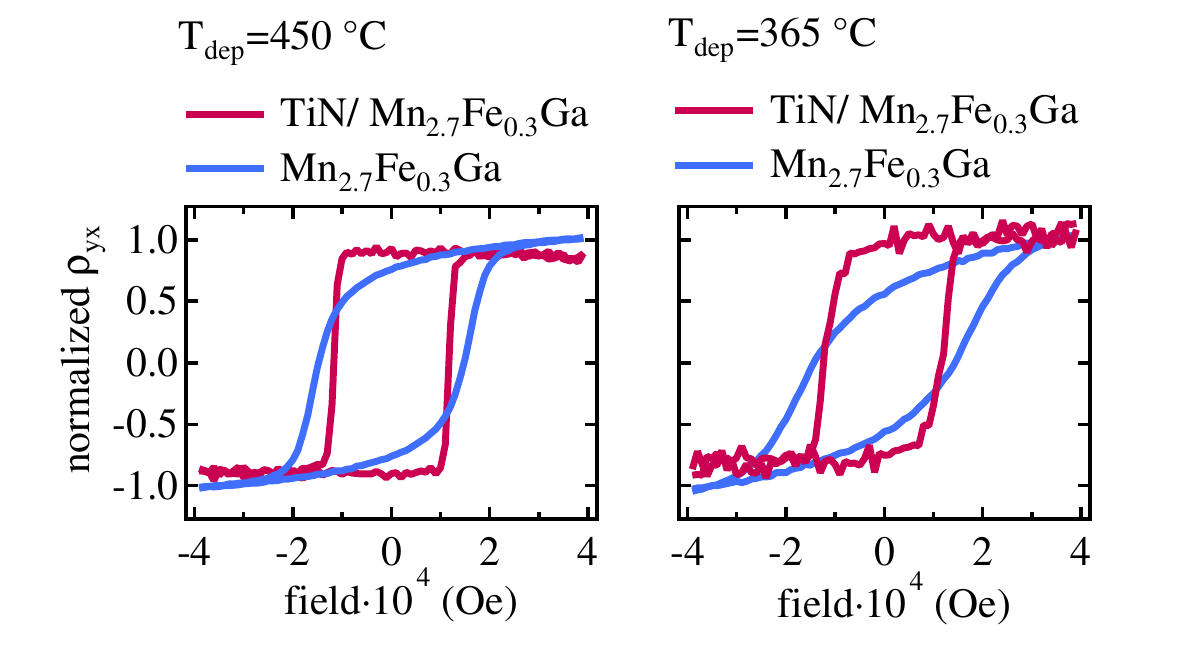}%
\caption{AHE hysteresis curves of \ce{TiN} buffered (red) and unbuffered (blue) \ce{Mn}-\ce{Fe}-\ce{Ga} measured at room temperature. Both samples were deposited on \ce{MgO} substrates. The \ce{TiN} seed-layer (\SI{30}{\nano\m} thickness) was deposited at \SI{405}{\degreeCelsius}. The \ce{Mn_{2.7}Fe_{0.3}Ga} thin films were deposited at \SI{450}{\degreeCelsius} and \SI{365}{\degreeCelsius}, respectively.}%
\label{fig:AHE_woTiN}%
\end{figure}

%
The magnetic properties were investigated via AGM (alternating gradient magnetometry) and AHE (anomalous Hall effect) measurements. The coercivity of the \ce{Mn}-\ce{Fe}-\ce{Ga} thin films was determined via AHE measurements in a 4-terminal arrangement, carried out in a closed cycled \ce{He}-cryostat. 
High coercivity fields (\(H_{c} \leq\) \SI{2e4}{}\,\rm{Oe}) in the oop direction can be reached by tuning the composition. The coercivity decreases with increasing \ce{Fe} and decreasing \ce{Mn} amount for both substrate types. The highest coercivity of \SI{1.8e4}{}\,\rm{Oe} was measured for \ce{Mn_{3}Fe_{0.4}Ga}. The lowest value of \SI{0.2e4}{}\,\rm{Oe} showed the cubic \ce{Mn_{1.3}Fe_{1.4}Ga} (Fig. \ref{fig:Hc_vs_Fe_T}).
Each sample revealed a hard magnetic axis in the in-plane direction. 
Since the samples could not be saturated in the in-plane direction even at an applied field of \SI{4e4}{}\,\rm{Oe}, a strong perpendicular magnetocrystalline anisotropy (PMA) could be experimentally verified. 
Fig. \ref{fig:AHE} shows a comparison of normalized magnetization \(M\) and AHE curves (\(U_{H}\)) for the tetragonally distorted \ce{Mn}-\ce{Fe}-\ce{Ga} deposited on \ce{STO} (smooth films with roughness values below \SI{1}{\nano\m}). In case of the \ce{Mn_{1.2}Fe_{2.6}Ga} and \ce{Mn_{1.5}Fe_{0.9}Ga} deposited on \ce{STO} substrates the hysteresis curves are in good agreement. The coercivity, the squareness ratio and the saturation magnetization values are given in Tab. \ref{TAB1:Hc}. We defined the squareness as \(S_{R}=M_{r}/M_{s}\) or \(S_{R}={\rho_{yx}}_{r}/{\rho_{yx}}_{s}\) (the \(r\) or \(s\) index denotes the remanence or the saturation value of the magnetization \(M\) or resistivity \(\rho\)). Increasing the \ce{Mn} and lowering the amount of \ce{Fe} obviously also leads to an increase of the squareness ratio. The magnetization values, determined via AGM measurements, are in the range of \SI{140}{\kilo\ampere/\m} and \SI{350}{\kilo\ampere/\m}, which is comparable to the values of the \ce{Mn}-\ce{Ga} compound.\cite{Glas:2013vc}
\ce{Mn_{1.6}Fe_{0.3}Ga} shows a feature in the AGM curves around \num{0}\,\rm{Oe} field, which could not be observed via AHE measurements. This was attributed to a second phase (soft magnetic) inside the \ce{Mn}-\ce{Fe}-\ce{Ga} thin film, which has a different coercive field. Such behavior was already observed for \ce{Mn_{2}Fe_{1}Ga} samples.\cite{Gasi:2013fe} The AHE measurements do not show such a feature, which indicates that the second phase could have a high resistance and therefore does not contribute to the AHE. A second phase was not detected via XRD measurements, leading to the assumption that this phase is an amourphous part of the material, which might be located at the grain boundaries.
To increase the applicability of \ce{Mn}-\ce{Fe}-\ce{Ga} as an electrode in MTJ's a \ce{TiN} seed-layer (\SI{30}{\nano\m} thickness) was used. Fig. \ref{fig:AHE_woTiN} shows a comparison of the AHE mesurements for \ce{TiN} buffered \ce{Mn_{2.7}Fe_{0.3}Ga} deposited at two different temperatures. In both cases the \ce{TiN} buffered samples revealed sharper switching of the magnetization compared to the unbuffered layers. The squareness ratio increases and the coercive field decreases for the buffered samples (Tab. \ref{TAB2:TiN}). Regarding the similar morphology of the samples, the reason for the changed switching behavior is still unclear and we attempt further investigations on this topic.
\begin{table}
\centering
\includegraphics[width=3.5in]{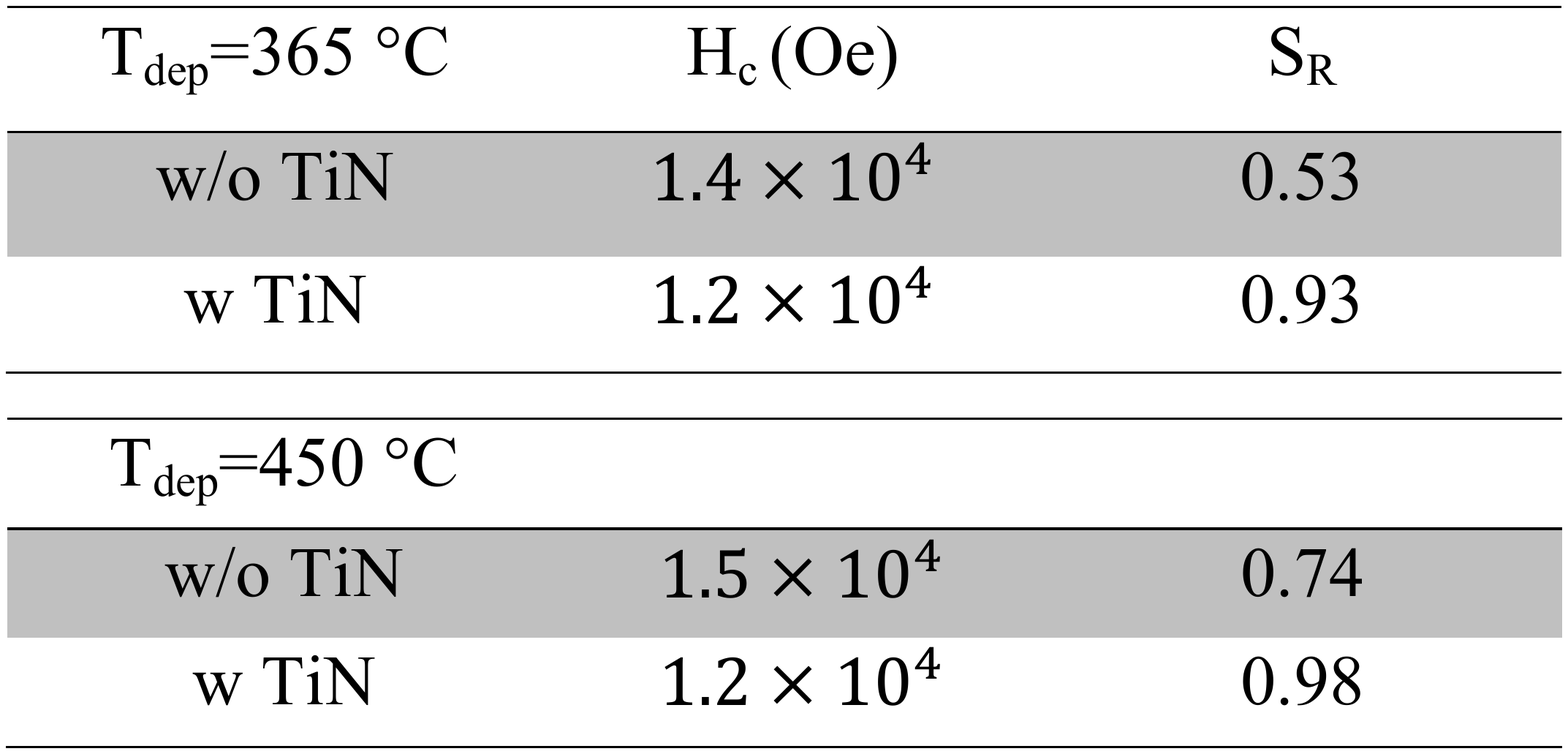}
\vspace{-3cm}
\caption{Comparison of the coercivity and squareness ratio values of \ce{TiN} buffered (w \ce{TiN}) and unbuffered (w/o \ce{TiN}) \ce{Mn_{2.7}Fe_{0.3}Ga} deposited on \ce{MgO} substrates. }
\label{TAB2:TiN}
\end{table}

\section{Conclusion}

Structural and  magnetic properties of sputter deposited \ce{Mn}-\ce{Fe}-\ce{Ga} compounds were investigated. Depending on the stoichiometry, the deposition temperature, as well as the used substrate (STO (001) and MgO (001)) or buffer layer (TiN) the \ce{Mn}-\ce{Fe}-\ce{Ga} crystallizes in the cubic \(D0_{3}\) (\(c\) = \SI{6}{\angstrom}) or the tetragonally distorted phase \(D0_{22}\) (\(c\) = \SI{7.15}{\angstrom}). The main drawback for applications is the island growth, which was confirmed via  AFM measurements. Low roughness (\(\leq \SI{1}{\nano\meter}\)) and the \(D0_{22}\) phase was observed for each used deposition temperature and composition of the \ce{Mn}-\ce{Fe}-\ce{Ga} on \ce{STO}, whereas on \ce{MgO} substrates the roughness showed strong dependence on the deposition temperature and the \ce{Mn}-\ce{Fe}-\ce{Ga} composition. Strong PMA was confirmed via AHE and AGM measurements. 
High coercivity fields (up to \(H_{c} \leq\) \num{2} \(\cdot 10^{4}\)\,\rm{Oe}) in the out-of-plane direction were reached by tuning the composition. \ce{TiN} buffered \ce{Mn_{2.7}Fe_{0.3}Ga} revealed sharper switching of the magnetization compared to the unbuffered layers. Similar results were achieved for \ce{TiN} buffered \ce{Mn}-\ce{Ga} and \ce{Co_{2}FeAl} thin films.\cite{Niesen2015}

\section*{Acknowledgments}
The authors gratefully acknowledge financial support by the Deutsche Forschungsgemeinschaft (DFG, Contract No. RE 1052/32-1).

\ifCLASSOPTIONcaptionsoff
  \newpage
\fi

\appendices
\section{}
The roughness values (root mean square) were calculated using a standard deviation of surface heights: 
\begin{equation}
 RMS=\sqrt{\frac{1}{N}\Sigma^{N}_{x=1} (z(x,y)-\overline z(N,M))^2}
\end{equation}
with \(\overline z(N,M)\) the arithmetic average height. The surface is described by a matrix with N lines and M columns corresponding to the points (x,y) of the height z(x,y). 


\bibliographystyle{IEEEtran}
\bibliography{bibfile}

\begin{thebibliography}{10}
\providecommand{\url}[1]{#1}
\csname url@samestyle\endcsname
\providecommand{\newblock}{\relax}
\providecommand{\bibinfo}[2]{#2}
\providecommand{\BIBentrySTDinterwordspacing}{\spaceskip=0pt\relax}
\providecommand{\BIBentryALTinterwordstretchfactor}{4}
\providecommand{\BIBentryALTinterwordspacing}{\spaceskip=\fontdimen2\font plus
\BIBentryALTinterwordstretchfactor\fontdimen3\font minus
  \fontdimen4\font\relax}
\providecommand{\BIBforeignlanguage}[2]{{%
\expandafter\ifx\csname l@#1\endcsname\relax
\typeout{** WARNING: IEEEtran.bst: No hyphenation pattern has been}%
\typeout{** loaded for the language `#1'. Using the pattern for}%
\typeout{** the default language instead.}%
\else
\language=\csname l@#1\endcsname
\fi
#2}}
\providecommand{\BIBdecl}{\relax}
\BIBdecl

\bibitem{Wolf:2001fu}
S.~A. Wolf, \emph{Science}, vol. 294, no. 5546, p. 1488, 2001.

\bibitem{Reiss2007}
G.~Reiss and D.~Meyners, \emph{\BIBforeignlanguage{English}{Journal of Physics:
  Condensed Matter}}, vol.~19, no.~16, p. 165220, 2007.

\bibitem{Thomas2006}
A.~Thomas, D.~Meyners, D.~Ebke, N.-N. Liu, M.~D. Sacher, J.~Schmalhorst,
  G.~Reiss, H.~Ebert, and A.~H{\"u}tten, \emph{Applied Physics Letters},
  vol.~89, no.~1, p. 012502, 2006.

\bibitem{Balke:2007eb}
B.~Balke, G.~H. Fecher, J.~Winterlik, and C.~Felser, \emph{Applied Physics
  Letters}, vol.~90, no.~15, p. 152504, 2007.

\bibitem{Winterlik:2008fj}
J.~Winterlik, B.~Balke, G.~Fecher, C.~Felser, M.~Alves, F.~Bernardi, and
  J.~Morais, \emph{Physical Review B}, vol.~77, no.~5, p.~12, 2008.

\bibitem{Wurmehl:2006db}
S.~Wurmehl, H.~C. Kandpal, G.~H. Fecher, and C.~Felser, \emph{Journal of
  Physics: Condensed Matter}, vol.~18, no.~27, p. 6171, 2006.

\bibitem{Glas:2013we}
M.~Glas, C.~Sterwerf, J.-M. Schmalhorst, D.~Ebke, C.~Jenkins, E.~Arenholz, and
  G.~Reiss, \emph{Journal of Applied Physics}, vol. 114, no.~18, p. 183910,
  2013.

\bibitem{Winterlik:2012cp}
J.~Winterlik, S.~Chadov, A.~Gupta, V.~Alijani, T.~Gasi, K.~Filsinger, B.~Balke,
  G.~H. Fecher, C.~A. Jenkins, F.~Casper, J.~K{\"u}bler, G.-D. Liu, L.~Gao,
  S.~S.~P. Parkin, and C.~Felser, vol.~24, no.~47, p. 6283, 2012.

\bibitem{Wollmann:2014ke}
L.~Wollmann, S.~Chadov, J.~K{\"u}bler, and C.~Felser, \emph{Physical Review B},
  vol.~90, no.~21, p. 214420, 2014.

\bibitem{Felser:kh}
C.~Felser, V.~Alijani, J.~Winterlik, S.~Chadov, and A.~K. Nayak, \emph{IEEE
  Transactions on Magnetics}, vol.~49, no.~2, p. 682, 2013.

\bibitem{Magnus:2011wl}
F.~Magnus, A.~S. Ingason, S.~Olafsson, and J.~T. Gudmundsson, \emph{Thin Solid
  Films}, 2011.

\bibitem{Krockenberger:2012bo}
Y.~Krockenberger, S.-i. Karimoto, H.~Yamamoto, and K.~Semba, \emph{Journal of
  Applied Physics}, vol. 112, no.~8, p. 083920, 2012.

\bibitem{Pritschow:07}
M.~Pritschow, ``Titannitrid- und titan-schichten f{\"u}r die
  nano-elektromechanik,'' Ph.D. dissertation, Institut f{\"u}r Mikroelektronik
  Stuttgart, Mechanical Engineering Department, 2007.

\bibitem{Niesen2015}
A.~Niesen, M.~Glas, J.~Ludwig, J.-M. Schmalhorst, R.~Sahoo, D.~Ebke,
  E.~Arenholz, and G.~Reiss, \emph{Journal of Applied Physics}, vol. 118,
  no.~24, p. 243904, 2015.

\bibitem{Glas:2013vc}
M.~Glas, D.~Ebke, I.~M. Imort, P.~Thomas, and G.~Reiss, \emph{Journal of
  Magnetism and Magnetic Materials 333}, p. 134, 2013.

\bibitem{Gasi:2013fe}
T.~Gasi, A.~K. Nayak, J.~Winterlik, V.~Ksenofontov, P.~Adler, M.~Nicklas, and
  C.~Felser, \emph{Applied Physics Letters}, vol. 102, no. 202402, 2013.

\end{thebibliography}

\end{document}